\documentclass[twocolumn,prl,showpacs,preprintnumbers,amsmath,amssymb,superscriptaddress,10pt]{revtex4-1}
\usepackage{amsmath,amssymb,amsfonts} 	
\usepackage{graphicx}
\usepackage[english]{babel}
\usepackage[utf8]{inputenc}
\usepackage{subfigure}
\usepackage{xcolor}
\usepackage[normalem]{ulem}
\usepackage{SIunits}
\usepackage[colorlinks=true,linkcolor=blue,citecolor=blue,urlcolor=black]{hyperref}
\usepackage{multirow}
\makeatletter
\def\bbl@set@language#1{%
  \edef\languagename{%
    \ifnum\escapechar=\expandafter`\string#1\@empty
    \else\string#1\@empty\fi}%
  \@ifundefined{babel@language@alias@\languagename}{}{%
    \edef\languagename{\@nameuse{babel@language@alias@\languagename}}%
  }%
  \select@language{\languagename}%
  \expandafter\ifx\csname date\languagename\endcsname\relax\else
    \if@filesw
      \protected@write\@auxout{}{\string\select@language{\languagename}}%
      \bbl@for\bbl@tempa\BabelContentsFiles{%
        \addtocontents{\bbl@tempa}{\xstring\select@language{\languagename}}}%
      \bbl@usehooks{write}{}%
    \fi
  \fi}
\newcommand{\DeclareLanguageAlias}[2]{%
  \global\@namedef{babel@language@alias@#1}{#2}%
}
\makeatother

\DeclareLanguageAlias{en}{english}

\renewcommand{\[}{\begin{equation}}
\renewcommand{\]}{\end{equation}}
\def\bea{\begin{eqnarray}}
\def\eea{\end{eqnarray}}

\def\runtime{(\the\time)\qquad\the\month/\the\day/\the\year}
\def\today
 {\count10=\year\advance\count10 by -2000 \number\day--\ifcase
  \month \or Jan\or Feb\or Mar\or Apr\or May\or Jun\or
             Jul\or Aug\or Sep\or Oct\or Nov\or Dec\fi--\number\count10}

\def\hour{\count10=\time\count11=\count10
\divide\count10 by 60 \count12=\count10
\multiply\count12 by 60 \advance\count11 by -\count12\count12=0
\number\count10 :\ifnum\count11 < 10 \number\count12\fi\number\count11}

\newcommand\THEOSMARVEL{Theory and Simulation of Materials (THEOS) and National Centre for Computational Design and Discovery of Novel Materials (MARVEL), {\'E}cole Polytechnique F{\'e}d{\'e}rale de Lausanne, 1015, Switzerland}
\newcommand{\dqmp}{Department of Quantum Matter Physics, University of Geneva, 24 Quai Ernest Ansermet, CH-1211 Geneva, Switzerland}
\begin{document}
\title{Abundance of $\mathbb{Z}_2$ topological order in exfoliable two-dimensional insulators}
\author{Antimo Marrazzo}
\email{antimo.marrazzo@epfl.ch}
\affiliation{\THEOSMARVEL}
\author{Marco Gibertini}
\affiliation{\THEOSMARVEL}
\affiliation{\dqmp}
\author{Davide Campi}
\affiliation{\THEOSMARVEL}
\author{Nicolas Mounet}
\affiliation{\THEOSMARVEL}
\author{Nicola Marzari}
\email{nicola.marzari@epfl.ch}
\affiliation{\THEOSMARVEL}

\date{\today}

\begin{abstract}
Quantum spin Hall insulators are a class of two-dimensional materials with a finite electronic band gap in the bulk and gapless helical edge states. In the presence of time-reversal symmetry,  $\mathbb{Z}_2$ topological order distinguishes the topological phase from the ordinary insulating one. Some of the phenomena that can be hosted in these materials, from one-dimensional low-dissipation electronic transport to spin filtering, could be very promising for many technological applications in the fields of electronics, spintronics and topological quantum computing. Nevertheless, the rarity of two-dimensional materials that can exhibit non-trivial $\mathbb{Z}_2$ topological order at room temperature hinders development. Here, we screen a comprehensive database we recently created of 1825 monolayers that can be exfoliated from experimentally known compounds, to search for novel quantum spin Hall insulators. Using density-functional and many-body perturbation theory simulations, we identify 13 monolayers that are candidates for quantum spin Hall insulators, including high-performing materials such as AsCuLi$_2$ and jacutingaite (Pt$_2$HgSe$_3$). We also identify monolayer Pd$_2$HgSe$_3$ as a novel Kane-Mele quantum spin Hall insulator, and compare it with jacutingaite. Such a handful of promising materials are mechanically stable and exhibit $\mathbb{Z}_2$ topological order, either unpertubed or driven by a small amount of strain. Such screening highlights a relative abundance of $\mathbb{Z}_2$ topological order of around 1\%, and provides an optimal set of candidates for experimental efforts.
\end{abstract}

\date{run through \LaTeX\ on \today\ at \hour}
\maketitle\bigskip \bigskip
\section{Introduction}
Since the very first discovery of the quantum spin Hall insulating state in HgTe quantum wells in 2007 \cite{hgte_mol_06}, the study of topological states of matter has seen enormous progress both in theory and experiments \cite{zhang_review_17}. $\mathbb{Z}_2$  topological order, first discussed by in Refs.~\cite{kane_quantum_2005,kane_z2_05} and \cite{bernevig_strain_06} for two-dimensional models, has also been formalised for three-dimensional materials \cite{fu_topo3d_07}, leading to the discovery of the so-called ``strong'' and ``weak'' topological insulators.
Although substantial progress has been made in predicting and confirming with experiments several three-dimensional topological insulators of different classes \cite{hasankane_review_10}, scant progress has been made in identifying their two-dimensional counterparts, i.e. quantum spin Hall insulators (QSHIs).  As of today, 1T'-WTe$_2$ is the only experimentally confirmed monolayer crystal hosting a robust QSHI phase up to 100 K \cite{wte2_sts_17,wte2_transp_17,100_science_18}; considering more general systems, Bi on a SiC  substrate stands out being a QSHI with a record-high measured band gap of 0.8 eV, owing to covalent bonding with the substrate \cite{reis_bismuthene_17}.

Besides being of fundamental scientific interest, several technological applications of topological insulators have been proposed. A broad class of applications is based on using the one-dimensional topologically protected states at the edge of a QSHI to realize low-dissipation nanowires, where the elastic backscattering is forbidden by time-reversal (TR) symmetry and electron transport is spin-momentum locked \cite{zhang_review_17}. For such applications, a large band gap would be beneficial not only to increase the operating temperature (limited by the intrinsic semiconducting behaviour of the bulk) but also to increase the transverse localization length of the edge states \cite{bernevig_book_2013}. The latter could help to reduce inelastic backscattering with the bulk and, more relevantly, to suppress hybridisation effects between the two pairs of helical states at opposite edges of a ribbon that otherwise would gap the edge spectrum. In the so-called topological  field-effect transistor (TopoFET) \cite{qian_quantum_2014}, an out-of-plane applied electric field drives the system from a QSHI to a normal insulating phase and allows to switch on and off edge transport. It is clear that for TopoFETs applications, high-performance materials must exhibit an enhanced response to applied electric fields such that the topological phase transition occurs for realistic (i.e. sufficiently low) gate voltages. Coincidently, a larger band gap tipically implies a stronger QSHI phase and so larger critical fields for the topological phase transition. In addition, material-dependent effects,  such as thermal expansion or electron-phonon couplings would also drastically affect the operating temperature of the device \cite{antonius_temptopo_16,monserrat_temptopo_16}.   Hence, an accurate figure of merit for TopoFETs would include not only the zero-temperature band gap and band inversion, but also a number of other quantities such as the critical electric fields or temperature of the topological phase transition, the stability with respect to oxidation and the availability of good dielectric substrates. By looking at the current scenario it is clear that the known QSHI materials present challenges for devices and applications, although examples like Ref.~\cite{reis_bismuthene_17} (Bi on SiC) show that is realistic and plausible to substantially increase performance through a combination of novel materials and careful engineering. 

It is then compelling to seek for novel QSHI materials, possibly outperforming the state of the art. The optimal material might constitute a Pareto optimisation that satisfies several requirements including, but not restricted to, those mentioned above.
In this work, we use first-principles simulations to perform a systematic computational screening for QSHIs, using a combination of high-throughput density-functional theory (DFT) calculations, to first identify interesting candidates, and density-functional pertubation theory (DFPT) and many-body perturbation theory at the G$_0$W$_0$ level (with spin-orbit coupling (SOC)) to provide accurate predictions on the most interesting materials. The results of this screening not only provide a distilled set of most promising QSHI candidates (several not reported before), that could focus experimental efforts, but it contributes to draw a comprehensive picture on the abudance and characterization of $\mathbb{Z}_2$ topological order in 2D materials.
\section{The screening protocol}
\begin{figure}[hbtp]
\centering
\includegraphics[width=0.5\textwidth]{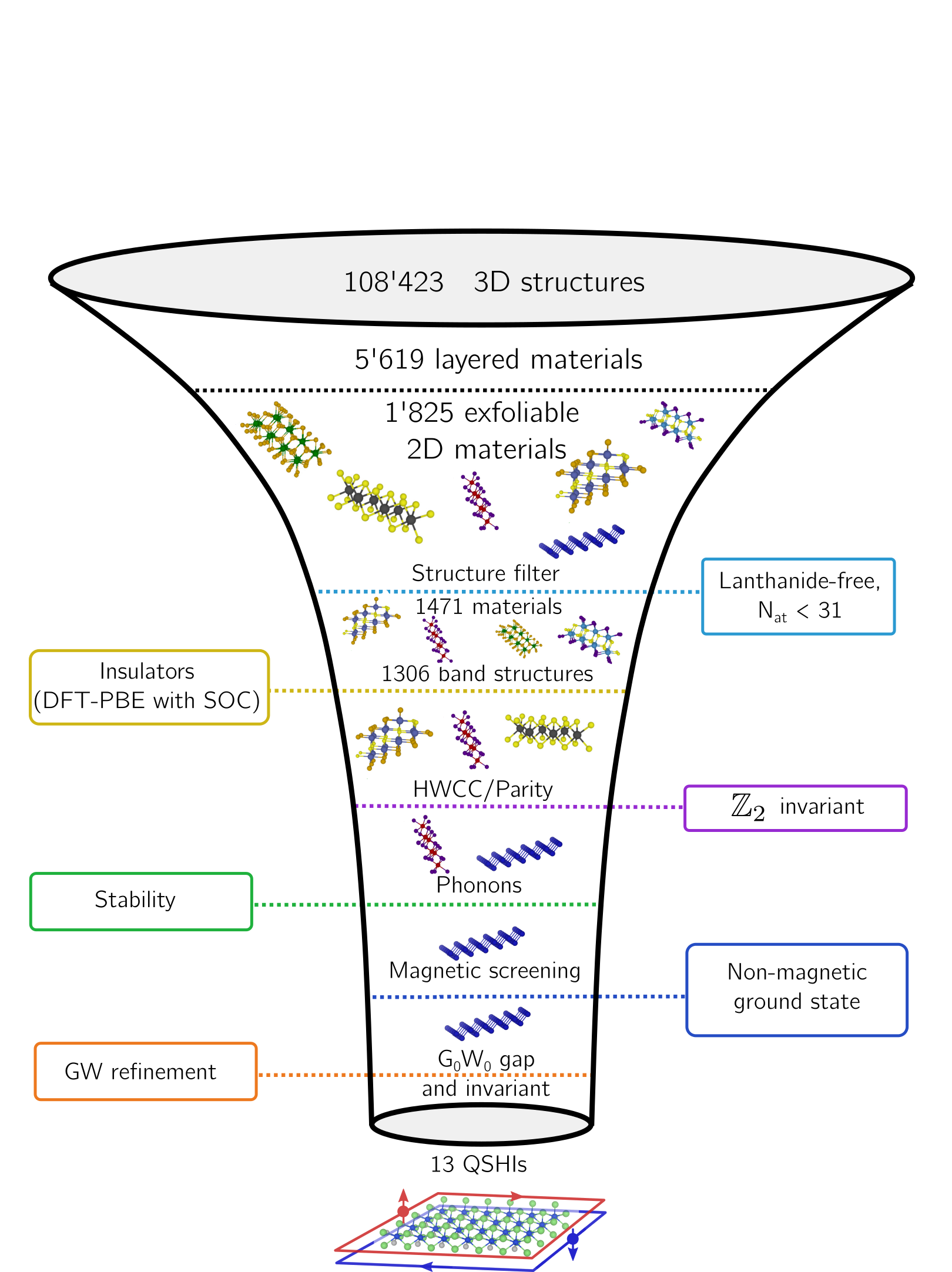}
\caption{\label{funnel}~Computational protocol for the screening of quantum spin Hall insulators (QSHIs) exfoliable from experimentally-known crystalline compounds. We start from the 1825 exfoliable 2D materials of Ref.~\cite{mounet_nanotech_18}, and each step reduces the number of potential candidates, with progressively more computationally-intensive calculations. The high-throughput approach based on density-functional theory calculations is complemented by density-functional perturbation theory and many-body perturbation theory calculations for the most interesting candidates. In particular, the topological phase is tested at the G$_0$W$_0$ level with spin-orbit coupling (SOC).}
\end{figure}

We systematically explore 1825 two-dimensional (2D) materials coming from experimentally known, exfoliable compounds, searching for QSHIs. All these 2D materials were identified in Ref.~\cite{mounet_nanotech_18} as easily or potentially exfoliable from their layered 3D parent crystals using extensive high-throughput first-principle calculations based on van-der-Waals DFT (vdW-DFT).  
Here, we select a set of computable properties defining good QSHIs candidates, and search the materials above in the hunt for novel QSHIs. In particular, we search for materials that are mechanically stable (i.e. whose phonons have no imaginary frequencies), that have a finite electronic band gap, a non-magnetic ground state and a non-trivial $\mathbb{Z}_2$ topological invariant. We adopt a computational high-throughput funnel approach (see Fig.~\ref{funnel}), where quantities that require less computational resources are computed first for the larger pool of structures, while more demanding properties are computed only for the progressively smaller sets of promising candidate materials. Given the manageable number of materials that we consider (of the order of thousands), we do not use approximate descriptors (as done earlier \cite{curtarolo_topoht_12,tavazza_spillagearxiv_18}) but we explicitly compute all properties from first principles.
To begin with, we optimize the monolayer geometry (cell vectors  and  atomic  positions) of the structures in set \cite{mounet_nanotech_18} with an upper limit on the number of atoms in the unit cell, both for their higher relevance and for computational efficiency.  So we restrict ourselves to all compounds  containing no more than 30 atoms in the unit cell, for a total number of 1582 structures. Then, lanthanides are removed, due to the limited accuracy of standard DFT in describing the electronic structures in such cases and the presence of multiple magnetic minima. The remaining compounds (1471 structures) are relaxed using the PBE functional assuming a non-magnetic ground state, yielding succesfully 1306 structures \cite{Note1}. At this point, we compute band structures along high-symmetry lines using DFT and the PBE functional with SOC, selecting all the band insulators \cite{Note2}.
We compute the $\mathbb{Z}_2$ invariant for all these by tracking the evolution of hermaphrodite Wannier charge centers (HWCC) \cite{sgiaro_herma_01,soluyanov_z2pack_11,z2pack_gresch_17}. 
At this point, all insulators with a non-trivial  $\mathbb{Z}_2$ invariant are considered tentative QSHIs and we assess their mechanical stability  by computing phonon dispersions (see Methods) using DFPT \cite{baroni_dfpt_review_01}.

Although some recent work has focused on substituting the calculations of topological invariants with the use of descriptors \cite{curtarolo_topoht_12,spillage_vanderbilt_14,tavazza_spillagearxiv_18} or on the elementary band representation \cite{bernevig_topoquantumchem_17}, we argue that the accuracy in the predictions of topological insulators is essentially driven by the accuracy of the calculated electronic structure and, in particular, of the spectral function. Substantial work has been made to find faster algorithms to compute the  $\mathbb{Z}_2$ and other invariants, henceforth facilitating the calculation of such invariants in large datasets of crystalline materials \cite{maya_nature_2019,tang_nature_2019,zhang_nature_2019} at a relatively low cost, essentially by maximally exploiting crystalline symmetries. However, for a reliable screening of topological insulators, most of the human and computational effort still goes into accurate band structure calculations (e.g. using G$_0$W$_0$ or dynamical mean field theory), in the study of possible magnetic ground states and in the assessment of mechanical stability (e.g. using phonons or molecular dynamics ) \cite{zunger_beware_2019}. 

   By their very nature topological insulators are characterised by topological invariants, i.e. by integers that do not change under smooth deformations. Such a feature is often used as an argument in favour of the robustness of topological properties, the most spectacular cases being the quantisation of the Hall conductance in the integer quantum Hall effect \cite{klitzing_rev_86} or the robust presence of topological insulator surface states with respect to surface reconstruction or different terminations. The issue is how easy it is, in practice, to perform a non-smooth deformation, which for a real material means to either break the protecting symmetries or close the gap. $\mathbb{Z}_2$ topological order is protected by time-reversal (TR) symmetry and so the system must be non magnetic: any time-reversal breaking perturbation would destroy topological order without even necessarily closing the gap. Hence, we select only materials whose ground state is non-magnetic based on the energetics of collinear DFT calculations for ferromagnetic and antiferromagnetic configurations (as in Ref.~\cite{mounet_nanotech_18}, see Methods). 

Then the question is whether DFT self-interaction, correlation effects, inaccuracies in the structural properties, or using Kohn-Sham (KS) states rather than Dyson orbitals are approximations severe enough to close the gap. For TR-invariant topological insulators such as QSHIs, the topological phase is driven by a combination of crystal symmetries, hybridisation and SOC, and each of these aspects must be accurately treated in order to get reliable predictions. Broadly speaking, the most significant approximation is using the KS states that underestimate the gap and often overestimate the strength of the band inversion, leading to possible false negatives (discarding as metals materials that are insulating) or false positive (predicting topological insulators that are actually trivial band insulators) \cite{zunger_false_2011}.
SOC is typically sufficiently well described by semilocal DFT, in particular when the band gap opening is driven by atomic SOC  \cite{oleg_bismuth_16} although quantitatively less accurately when SOC enters as a hopping term \cite{marrazzo_jacutingaite_18}.

Crystal symmetries are crucial and most predicted QSHIs actually fall in very few structure prototypes, such as the honeycomb lattice \cite{xenes_review_17} or the distorted 1T' phase  of transition-metal dichalcogenides \cite{qian_quantum_2014,qian_tellurides_17}. Recently, Ref.~\citep{bernevig_topoquantumchem_17} emphasised on the importance of site-symmetry groups, which impose strong constraints on the allowed connections between bands and hence determining the presence or absence of disconnected elementary band representations (i.e. topologically non-trivial manifolds) around the Fermi level. Hence it is very important to adopt a crystal structure that faithfully represents the experimental structure at least at low temperatures, where the vibrational effects can be neglected \cite{monserrat_phsplitting_17}. In this regard, it is preferable to start from experimental data even in the case of novel 2D materials that have not been studied in experiments, as the vast majority of known exfoliable 2D materials inherit the crystal symmetries of their parent 3D layered crystal and do not undergo structural transitions in the monolayer limit. 

Nonetheless,  Ref.~\cite{zunger_false_2011} highlighted how sensitive  $\mathbb{Z}_2$ topological order is with respect to the equilibrium lattice constant. In our protocol we always start from experimental crystal structures and perfom a first structural optimization using vdW-DFT; once identified as exfoliable, we perfom a second structural optimization for the monolayers using DFT-PBE. For materials that are identified as prospective QSHI candidates we perform an additional structural optimisation with tighther thresholds and more accurate, more computationally expensive, pseudopotentials (see Methods). Although we pay particular attention to structural optimizations, we also monitor deviations between the experimental and the calculated equilibrium lattice constants. Hence, we augment our protocol by considering all metals with a direct gap (DGM) along high-symmetry paths (built following Ref.~\cite{kpath_86}) and apply a $\pm 1,2,3 \%$ hydrostatic strain (see Methods). DGMs are chosen because they can be adiabatically connected to the insulating state and the $\mathbb{Z}_2$ invariant can be well defined considering the first $n_e$ bands where $n_e$ is the number of electrons, even if the Fermi level does not lie in a gap. If a DGM becomes an insulator under strain then the strained structure is added to the list of materials to be screened for QSHIs. This procedure has the two-fold purpose of identifying materials that would be QSHIs at their experimental lattice constant or that could be driven to be QSHI, e.g. by choosing a suitable substrate. There could still be systems that would not become insulators under strain at the DFT-PBE level and yet be insulators experimentally and recognised as such by e.g. many-body pertubation theory. Although such materials can exist and may even display robust band inversions, it is unlikely that they would have very large band gaps, hence we leave such investigation to future work. 

Finally, we address correlation effects by performing (very costly) many-body perturbation theory calculations (G$_0$W$_0$ with SOC) for the five most interesting QSHI candidates. These materials have been chosen as optimal in the sense they optimize the multidimensional requirements of low binding energy, large electronic band gap and strong band inversion. Some of them excel for specific aspects (e.g. large band gap) while keeping a sufficiently good performance on all the other relevant parameters, or because they show a good average over all parameters without being the top material in any of the quantities considered. 

Although the suppression of bulk transport is inherently related to the size of the global band gap, the robustness of the topological phases is mostly dictated by the magnitude of the band inversion. To be more precise, we define the \emph{inversion strength} (IS) for the two classes of QSHIs. For Bernevig-Hughes-Zhang (BHZ) QSHIs \cite{bernevig_science_06}, where there is a clear band inversion between electronic bands of different orbital character, the IS is defined as the energy difference between the lowest unoccupied and the highest occupied band at the high-symmetry point where the band inversion occurs (e.g. $\Gamma$ for Bi monolayers). For Kane-Mele QSHIs, where a single Dirac cone at K is gapped by the Kane-Mele SOC \cite{kane_quantum_2005,kane_z2_05}, we define the IS as the direct band gap at K. Hence we compute the inversion strength both at the DFT-PBE and G$_0$W$_0$ levels with SOC, by evaluating the direct gap at the relevant high-symmetry point. G$_0$W$_0$ calculations of 2D materials are known to be very challenging in terms of computational resources due to the strong dependence of the 2D dielectric function around $\mathbf{q}=0$ \cite{thygessen_gwconv_13,thygessen_2dgw_16}. Hence, we provide an accurate estimate of the band inversion by using a series of very dense $\mathbf{k}$-point grids (eg. up to $48\times48\times 1$ or more) and extrapolate to an infinitely dense grid (see Methods).
\\ 

\section{Results}

\begin{figure*}[hbtp]
\centering
\includegraphics[width=0.8\textwidth]{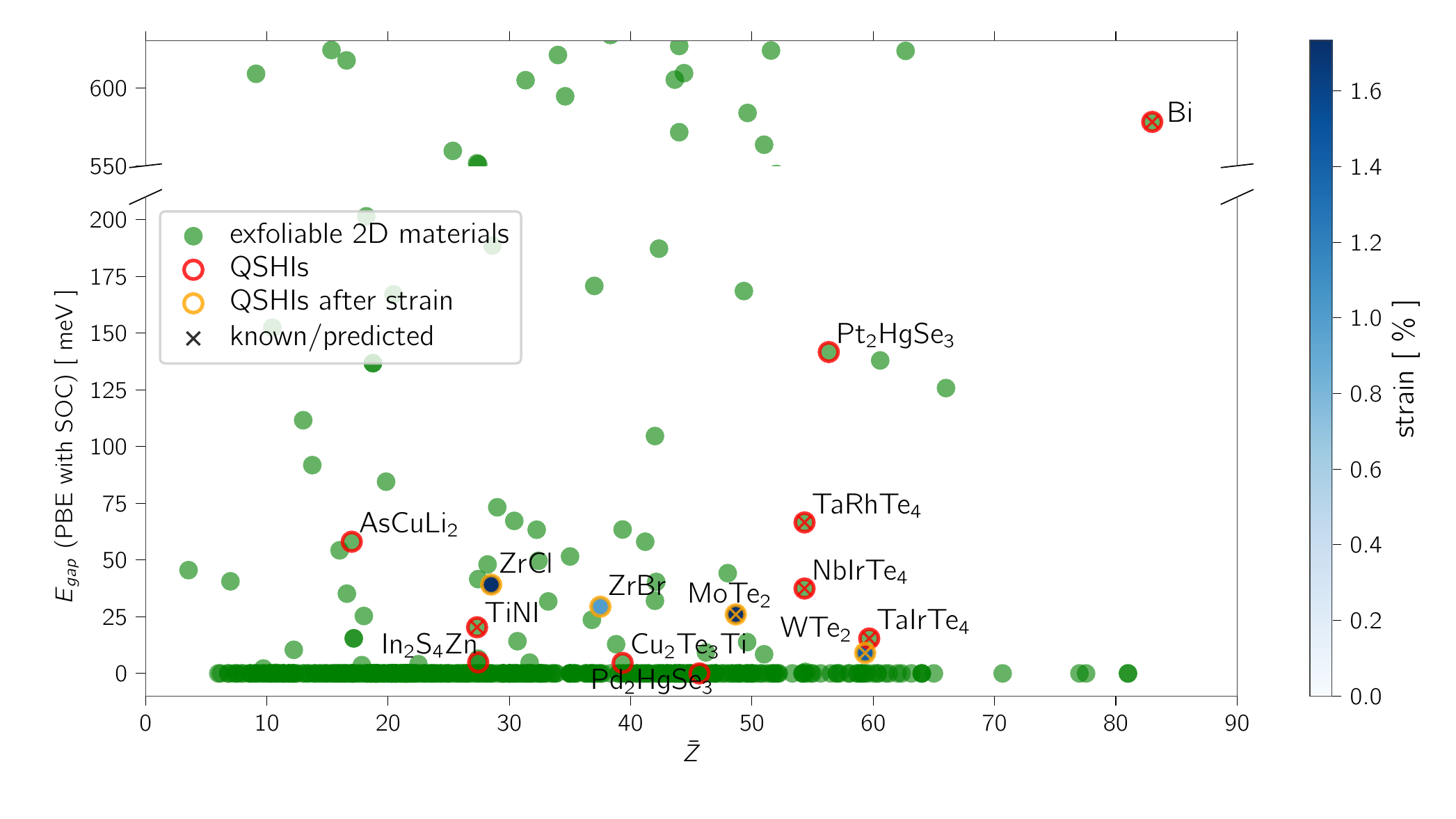}
\caption{\label{fig_scatter}Band gap at the DFT-PBE level with SOC versus average atomic number for exfoliable 2D materials identified in Ref.~\cite{mounet_nanotech_18}. QSHI candidates are denoted by red circles; strain-driven QSHIs are marked by orange circles and colored according to the minimum amount of strain to open a band gap. We identify 13 candidates (6 new) out of the 1306 screened, to which we add palladium jacutingaite (Pd$_2$HgSe$_3$), recently synthesised \cite{pd_vyz_2017} as the second ever KM QSHIs.}
\end{figure*}

\begin{table*}
\begin{tabular}{|c|c|c|c|c|c|c|c|c|}
\hline
Compound & Space & E$_b$  [meV$\cdot\angstrom^{-2}$] &  band gap  &  inversion strength & inversion strength & strain  & QSHI  &  Ref.\\
& group & (DF2-C09/rVV10) & PBE [meV]   & PBE   [meV]  & G$_0$W$_0$  [meV]  &  [$\%$] & type & \\
\hline
\hline
Bi & P${\bar{3}m1}$ (164) & 18/25 & 545 & 685 & 760
 & none  & BHZ  &\cite{mounet_nanotech_18}  \\
Pt$_2$HgSe$_3$ &  P${\bar{3}m1}$ (164) & 60/63 & 149 & 168 & 530 & none & KM  & \cite{marrazzo_jacutingaite_18} \\
Pd$_2$HgSe$_3$ &  P${\bar{3}m1}$ (164) & 61/66 & 0 & 80 &  41 & none & KM  & \cite{pd_vyz_2017} \\
TiNI & P${mmn}$(59) & 15/22 & 18 & 141 & -705 (trivial)&  none & BHZ & \cite{wang_tini_16,mounet_nanotech_18} \\
AsCuLi$_2$ &  P${\bar{6}m2}$ (187) & 63/62 &  45  & 80 &  169 &none & BHZ   &  \\
WTe$_2$ & P2$_{1}/m$ (11) &30/27   & 9 (2$\%$)  & 972 & 978 (from \cite{qian_quantum_2014}) & 2 $\%$& BHZ   & \cite{qian_quantum_2014,mounet_nanotech_18} \\
MoTe$_2$ & P2$_{1}/m$ (11)  &25/30 &  26 (3$\%$) & 408 & 403 (from \cite{qian_quantum_2014})  & 3 $\%$& BHZ   & \cite{qian_quantum_2014,mounet_nanotech_18}\\
TaIrTe$_4$ & P$m$ (6) & 26/31 & 11 & 204 &&none & BHZ  & \cite{qian_tellurides_17} \\
TaRhTe$_4$ &  P$m$ (6)  & 26/31 & 65 & 215 && none& BHZ  & \cite{qian_tellurides_17} \\
NbIrTe$_4$ &  P$m$ (6)  & 27/32 & 36 & 161 &&none & BHZ  & \cite{qian_tellurides_17} \\
Cu$_2$Te$_3$Ti &  C${2/m}$ (2) & 44/44 &  8    &21 & &none & BHZ   &  \\
In$_2$ZnS$_4$& P${3m1}$ (156) & 36/39 &   0   &191 & &none & BHZ   &  \\
ZrBr & P${\bar{3}m1}$ (164)  & 16/22 &  29 (1$\%$)  & 45 (1$\%$)  &  & 1 $\%$& BHZ   & \cite{dresden_htz2_2018}  \\
ZrCl &  P${\bar{3}m1}$  (164) & 15/22 & 39 (3$\%$)  & 60 (3$\%$)  &  & 3 $\%$& BHZ   &   \\
\hline
\end{tabular}
\caption{\label{table}QSHI candidates and their corresponding properties: chemical formula, space group, binding energy computed with two different van der Waals functional (DF2-C09 and rVV10, see Methods), band gap at the DFT-PBE level with SOC, inversion strength at the DFT-PBE level with SOC, inversion strength at the G$_0$W$_0$ level with SOC, miminum amount of strain to be insulating (DFT-PBE level), topological phase type (either Kane-Mele or Bernevig-Hughes-Zhang), reference (if present).}
\end{table*}
The results of the computational screening are summarized in Fig.~\ref{fig_scatter} and in Tab.~\ref{table}.
We find 13 QSHIs candidates using the protocol described above and they are listed in Tab. \ref{table}, together with key relevant properties: the space group, binding energy, band gap, inversion strength, underlying model (BHZ or KM), minimum amount of strain to drive an insulating topological phase and relevant references if present. In addition, we present in the Supporting Information the crystal structures, the DFT-PBE band structures (with and without SOC), and the DFPT phonon dispersions (without SOC, except for jacutingaite) for each of these candidates. In Fig.~\ref{fig_scatter} we report the band gap at the DFT-PBE level versus the average atomic number of a given crystal structure for all the screened materials whose band gap is in a relevant energy range. All compounds are identified by a green disk, and QSHI candidates (at the DFT-PBE level) are marked with an additional red circle. In case the QSHI phase is driven by strain the structure is marked by an orange circle and the disk is colored in blue according to the minimum amount of strain that is necessary to drive a transition from the metallic to the insulating state (essentially opening an indirect band gap). 

\begin{figure*}[hbtp]
\centering
\includegraphics[width=0.45\textwidth]{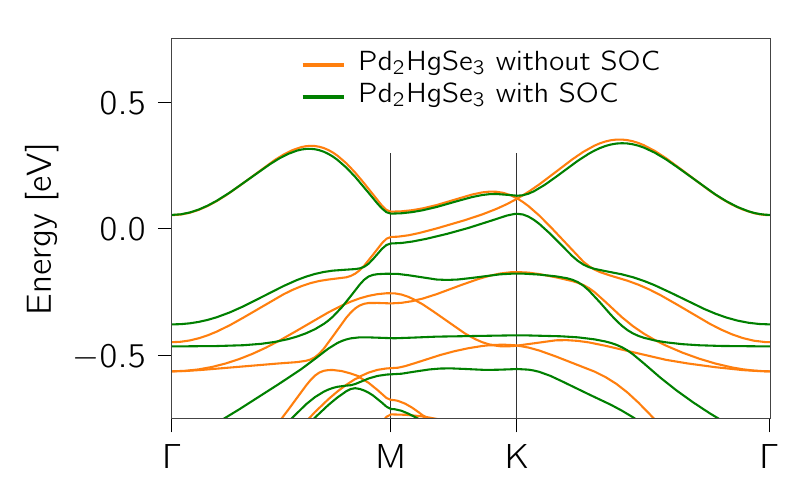}
\includegraphics[width=0.45\textwidth]{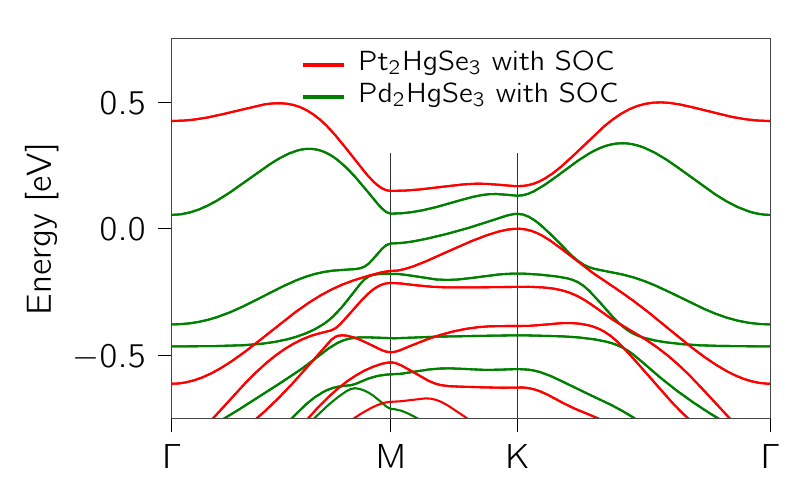}
\caption{\label{fig_pd_comp}Left panel: DFT-PBE band structure with (green) and without (orange) SOC for monolayer Pd$_2$HgSe$_3$, where a Dirac crossing at K is gapped by SOC, as in monolayer Pt$_2$HgSe$_3$ (jacutingaite). Right panel: DFT-PBE band structure with SOC for monolayer Pd$_2$HgSe$_3$ (green) and monolayer Pt$_2$HgSe$_3$ (red); the large band gap of the latter is driven by the presence of the heavy element Pt.}
\end{figure*}

\begin{figure*}[hbtp]
\centering
\includegraphics[width=0.4\textwidth]{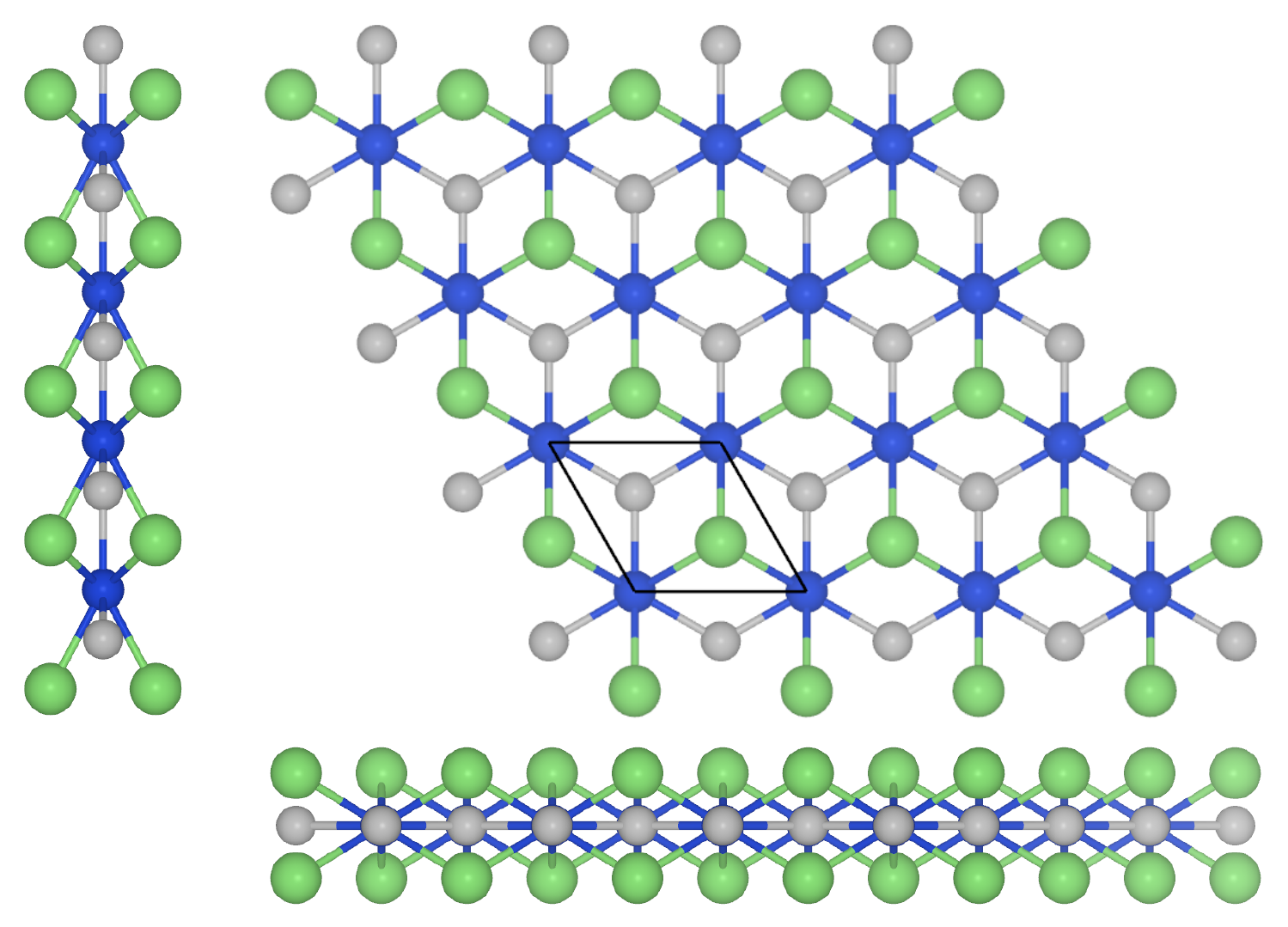}
\includegraphics[width=0.45\textwidth]{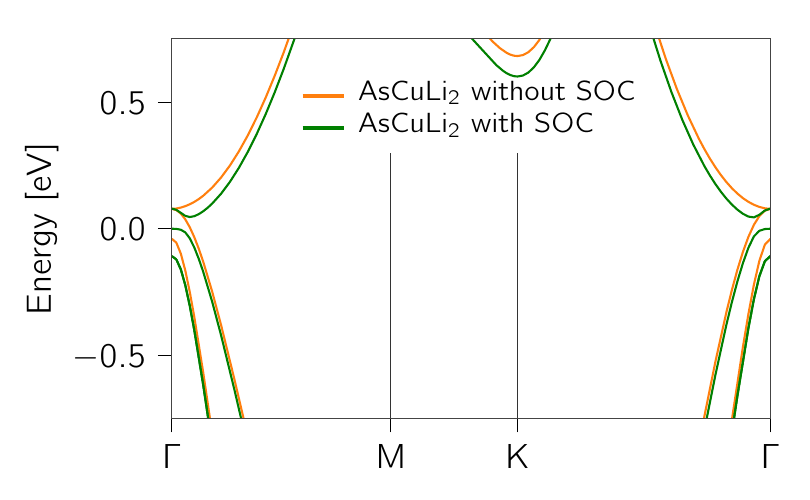}
\caption{\label{fig_asculi2}Left panel:  Crystal structure of AsCuLi$_2$ (top and lateral views), a potentially exfoliable material identified in Ref.~\cite{mounet_nanotech_18}. Right panel: DFT-PBE band structure with (green) and without (orange) SOC for monolayer AsCuLi$_2$, where SOC drives a band inversion at $\Gamma$. }
\end{figure*}

As expected all candidates are essentially narrow gap semiconductors, with band gaps lower than 0.6 eV driven by SOC. A part from the obvious observation that SOC is due to the presence of heavy chemical elements, there is no strong correlation between the magnitude of the band gap and the highest or average atomic number for the atoms of a given structure. However, the two largest-gap candidates, namely Bi and Pt$_2$HgSe$_3$, indeed contain three of the heaviest non-radioactive elements of the periodic table (Pt, Hg, Bi). 
In addition to the exfoliable materials of Ref.~\cite{mounet_nanotech_18}, we added Pd$_2$HgSe$_3$ which was not included in ICSD or COD at the time the study of Ref.~\cite{mounet_nanotech_18} was performed.  Pd$_2$HgSe$_3$ is a crystalline compound that has recently been identified experimentally \cite{pd_vyz_2017,pdphasediag_vyz_2014} and whose structure is identical to jacutingaite (Pt$_2$HgSe$_3$) and is also potentially exfoliable with a very similar binding energy ($\sim$60 meV$\cdot\angstrom^{-2}$, see Tab. \ref{table}). 
We now first discuss individually the most interesting candidates and then comment on some trends.\\
\paragraph{Bi.}~Monolayer bismuth has long been predicted to be a QSHI \cite{bismuth2d_prl_06}, although experimental confirmation has proven to be difficult \cite{bismuthtopo_nphys_14,bionnb_prb_18, sabater_prl_13} and the isolation of a clean monolayer is still a challenge. With a very low binding energy ($\sim 20 \text{ meV}\cdot\angstrom^{-2}$), a strong band inversion (0.7 and about $0.8$ eV at the PBE and G$_0$W$_0$ level respectively) and very large band gap ($\sim 0.6$ eV with PBE), monolayer bismuth remains superior to all other candidates. Monolayer bismuth is a unary compound with a relative simple crystal structure, namely a buckled honeycomb lattice with two atoms per unit cell, that makes it appealing also from the point of view of experimental synthesis. Indeed, a recent experimental effort reported that Bi on a SiC substrate is a record-high QSHI with a 0.8 eV band gap, although the covalent bonding substantially alters the atomic and electronic structure of the monolayer \cite{reis_bismuthene_17}.
\paragraph{Pt$_2$HgSe$_3$/Pd$_2$HgSe$_3$.}~Jacutingaite (Pt$_2$HgSe$_3$) and Pd$_2$HgSe$_3$ share very similar band structures (see Fig.~\ref{fig_pd_comp}). They both realize the Kane-Mele model with a Dirac cone at K split by SOC, although they also exhibit some differences. Pd$_2$HgSe$_3$ is metallic at the level of DFT-PBE: although SOC opens a gap at K, the bottom of the conduction band (at $\Gamma$) is degenerate with the top of the valence band (at K), realizing a peculiar semimetal with a finite direct gap at each $\mathbf{k}-$point. G$_0$W$_0$  quasiparticle corrections open a gap and show that monolayer Pd$_2$HgSe$_3$ is a Kane-Mele QSHI, with a band inversion estimated from extrapolations to be around $41$ meV.  In addition, the direct gap at K is roughly a third (0.07 eV with PBE) of the one of jacutingaite (0.17 eV with PBE); this is perfectly consistent with the analysis of Ref.~\cite{marrazzo_jacutingaite_18} where $2/3$ of the gap at K of jacutingaite is attributed to Pt and only a third to Hg.
We stress that these are the only two QSHI candidates of the Kane-Mele type we found (in addition to graphene itself which is not included here owing to its vanishing small band gap). 
\paragraph{TiNI.}~With a very low binding energy ($\sim 20 \text{ meV}\cdot\angstrom^{-2}$), a good inversion strength (0.17 eV with PBE) and sufficient band gap (0.03 eV with PBE), TiNI may seem a good candidate and indeed it has already been identified independently by different authors \cite{wang_tini_16,mounet_nanotech_18} as a QSHI. However G$_0$W$_0$ calculations point to TiNI actually being a trivial insulator  with a relatively large direct gap at $\Gamma$ of 0.7 eV. This highlights the fundamental importance of performing accurate quasiparticle calculations when predicting topological phases \cite{zunger_false_2011}, especially for 2D materials where dielectric screening is weak and the dielectric function has a strong spatial dependence \cite{thygessen_gwconv_13,thygessen_2dgw_16}. 
\paragraph{AsCuLi$_2$.}~This is a promising QSHI candidate, not reported before, with a good band inversion (0.08 eV with PBE) that gets actually stronger at the G$_0$W$_0$ level (0.17 eV) and a sufficient band gap (0.06 eV at the PBE level). The crystal and band structures of monolayer AsCuLi$_2$ are reported in Fig.~\ref{fig_asculi2}. The crystal structure can be decomposed into a honeycomb lattice made of As and Cu (as in hexagonal boron-nitride) sandwiched between two triangular lattices made of Li; the two Li sublattices sit above and below the mirror plane represented by the AsCu sublattice. This is a novel structural prototype for QSHIs, it has space group 187 and it is characterized by a clean band inversion at $\Gamma$. At the DFT-PBE level without SOC, the system is a filling-enforced semimetal, i.e. the Fermi surface displays nodal features at specific electron fillings that are protected by a combination of crystalline and time-reversal symmetries, characterised by a symmetry-protected four-fold degenerate point at $\Gamma$. The orbital character of the Fermi surface is a combination of $d_{x^2-y^2},d_{xy}$ orbitals of Cu and $p_x,p_y$ orbitals of As. Here SOC gaps the four-fold degeneracy point and opens an indirect gap, owing to the presence of the heavy element As.
\paragraph{W(Mo)Te$_2$.}~Transition metal dichalcogenids in the 1T' phase have been predicted to host robust QSHI phases \cite{xiang_quantum_2016}, with several experimental confirmations for the case of WTe$_2$  \cite{wte2_sts_17, wte2_transp_17, 100_science_18}. On the computational side, the strong band inversion of WTe$_2$ is present at different levels of theory, from PBE \cite{xiang_quantum_2016} to HSE \cite{wte_hse_16,wte2_sts_17} to G$_0$W$_0$ \cite{xiang_quantum_2016}. However, the evidence of a strictly insulating bulk with a finite band gap is extremely sensitive to the lattice constant and the techniques used to compute the band structure, where a finite gap appears with the HSE functional while no indirect gap is present with PBE and G$_0$W$_0$. At the PBE level, a small compressive strain ($2\%$) is enough to open a global gap. WTe$_2$ has a low binding energy ($\sim 30 \text{ meV}\cdot\angstrom^{-2}$) and it is known to be exfoliable with scotch-tape techniques \cite{100_science_18}. Among the QSHIs we find, WTe$_2$ currently considered the best monolayer QSHIs, is not studied in detail here as it has been thoroughly discussed in recent experimental and theoretical literature \cite{wte2_sts_17, wte2_transp_17, qian_quantum_2014, 100_science_18}.
\paragraph{MM'Te$_4$.}~Monolayers of the MM' class of tellurides, where M $=$ Nb, Ta and M' $=$ Ir, Rh,  had already been proposed in Ref.~\cite{qian_tellurides_17} and confirmed here to host the QSHI phase. Their low binding energy and good band inversion (up to $\sim 0.2$ eV in TaRhTe$_4$ ) makes them good candidates for experimental studies.

\paragraph{In$_2$ZnS$_4$.}~In addition to potentially novel high-performance QSHI candidates, the computational screening provides also novel ideas and prototypes that would possibly inspire the engineering of materials, with In$_2$ZnS$_4$ being an excellent example. In$_2$ZnS$_4$ is a potentially exfoliable material ($E_b \sim 35-40 \text{ meV}\cdot\angstrom^{-2}$) with a good band inversion and a vanishing small band gap, overall a not-so promising QSHI candidate. However, In$_2$ZnS$_4$ has a very peculiar crystal structure that could be thought of as being made of two separate 2D subunits, partially bonded together by van der Waals interactions. In fact, in order to properly refine the crystal structure we further optimize it by using a non-local vdW functional (DF2-C09 \cite{lee_df2_09, cooper_c09_10}, see Methods). The case of In$_2$ZnS$_4$ suggests the possibility of stacking monolayers with moderate interactions to engineer a topological phase, where the strength of the interactions is stronger than in simple vdW heterostructures and substantially weaker that in the case of Bi on a SiC substrate \cite{reis_bismuthene_17}. To this aim, one could explore all the possible combination of the $\sim$1000 potentially exfoliable materials identified in Ref.~\cite{mounet_nanotech_18}.

\paragraph{ZrBr(Cl).}~In these materials the QSHI phase, typically driven by crystal field and marked by relatively small band inversions with small or even vanishing indirect band gaps, is very sensitive to the lattice constant \cite{zunger_false_2011}. This can potentially be exploited by substrate engineering, where the choice of a suitable close-matched substrate can strain the monolayer and drive a QSHI phase. At the DFT-PBE level, ZrBr has a small band inversion and it is just on the edge of being an insulator, with an indirect gap that opens already at 1$\%$ of isotropic strain. ZrCl has the same prototype of ZrBr, although it requires a little bit more strain (3$\%$) to open a band gap.

Although the limited number (13) of QSHI candidates does not allow to use statistics, it is worth mentioning that one-third of all candidates have a crystal structure with space group  P${\bar{3}m1}$  (164) and all candidates have less than 12 atoms per unit cell (our search was conducted on all exfoliable materials with up to 30 atoms per unit cell). This hints at a possible correlation between the presence of a QSHI phase and particular structural motifs or specific point groups in small crystal structures, and it will be studied in future work. In addition, the presence of very heavy elements such Bi, Hg, Pt (and probably Pb \cite{olsen_z2ht_2019}) seems to be beneficial for large band gaps, independently of the precise mechanisms through which SOC opens a band gap. 
\section{Conclusions}
Using a combination of high-throughput DFT techniques and accurate DFPT and MBPT-G$_0$W$_0$ calculations,  we have screened 1306 out of 1825 exfoliable two-dimensional materials proposed in Ref.~\cite{mounet_nanotech_18}, to search for novel QSHIs. We have identified 13 monolayers as dynamically stable and potentially exfoliable QSHIs.  By using the high-throughput protocol described here, we identify several  compounds that have already been predicted as QSHIs (e.g. TaRhTe$_4$ or TiNI) and, in a few cases, also confirmed with experiments (Bi, WTe$_2$), validating the approach. In addition, we found several novel candidates, including promising high-performance materials such as AsCuLi$_2$, a second Kane-Mele QSHI Pd$_2$HgSe$_3$ or inspiring novel prototypes such In$_2$ZnS$_4$. 
We also showed the importance of adopting accurate theoretical frameworks to deal with SOC and excited states, which can dramatically affect the prediction of topological phases; this is most remarkable in the case of TiNI, predicted to be a QSHI by DFT-PBE calculations and found here to be trivial at the level of G$_0$W$_0$.
This screening effort points to a relative abundance of $\mathbb{Z}_2$ topological order in two-dimensional insulators of around 1$\%$. We want to remark that this does not imply that topological order is a rare property; at least for 3D materials this has been shown not to be the case \cite{bernevig_topoquantumchem_17}, with a sizeable fraction of all crystals showing electronic manifolds that are topologically non-trivial in a broad sense  \cite{maya_nature_2019,tang_nature_2019,zhang_nature_2019}.
What seems to be relatively rare is the simultaneous occurrence of a true 2D bulk insulating phase that exhibits $\mathbb{Z}_2$ topological order considering the entire manifold of the occupied valence bands, which phenomenologically coincides with the presence of strictly 1D electronic transport dominated by topologically-protected helical edge states. We differentiate such systems, both topological and truly insulating, from the general case of metals with a well defined interband gap where non-trivial topological invariants can be well defined. In the latter case, topologically-protected gapless edge states are superimposed to the gapless bulk energy spectrum; the entire system, bulk and terminations alike, behave as a metal. As a matter of fact, only systems where the electron transport happens exclusively at the 1D topologically-protected helical edge states can be of interest for devices and applications, allowing to exploit the absence of elastic backscattering, spin-momentum locking and robustness to pertubations.
 
The present screening provides a useful set of promising and novel QSHI candidates that would ideally prompt further experimental efforts. Finally, we highlight that extensive computational materials screening of this kind provide a unique advantage of finding unexpected novel prototypes and mechanisms that do not fit into the existing knowledge and could possibly inspire, as in the case of In$_2$ZnS$_4$, novel engineering strategies.

\section{Acknowledgements} 
This  work  was  supported  by  the  NCCR  MARVEL  of  the  Swiss  National  Science  Foundation.   We acknowledge PRACE for awarding us access to MARCONI hosted at CINECA, Italy and Piz Daint hosted at CSCS, Switzerland. Simulation time on  Piz Daint was also provided by CSCS though the production proposal s580. M.G.\ acknowledges support from the Swiss National Science Foundation through the Ambizione program. 
\section{Methods}
DFT calculations are performed with the Quantum ESPRESSO distribution \cite{qe_paper_09,qe_paper_17}, using the PBE functional and the SSSP \cite{sssp_arxiv_18}  and PseudoDojo \cite{hamann_oncv_13,dojo_paper_18} pseudopotentials’ libraries. The SSSP library \cite{sssp_arxiv_18} contains two sets of extensively tested pseudopotentials from various sources: the SSSP efficiency, tailored for high-throughput materials screening, and the SSSP precision, developed for high-precision materials modelling. As of today, the SSSP precision library is the most precise open-source pseudopotential library available \cite{sssp_arxiv_18} compared to all-electron reference data \cite{cotto_science_16}.
For the SSSP library wavefunction and charge-density cutoffs are chosen according to convergence tests with 
respect to cohesive energy, pressure, band structure and phonon frequencies performed for each individual element, as discussed in Ref.~\cite{sssp_arxiv_18}.  For the fully-relativistic PseudoDojo library \cite{dojo_paper_18} wavefunction cutoffs are chosen according to convergence tests with respect to the band strucure of elemental crystals, in particular converging the $\eta_v$  and $\eta_{10}$ below 10 meV and $\text{max } \eta_v$  and  the $\text{max } \eta_{10}$ below 15 meV respectively (see Ref.~\cite{sssp_arxiv_18}).  For the high-throughput calculations without (with) SOC we use a $\mathbf{k}$-point density of 0.2 $\angstrom^{-1}$ using a Marzari-Vanderbilt smearing \cite{mv_smearing_99} of 0.02 Ry and the SSSP efficiency v1.0 (PseudoDojo) library. Structural optimization, band structures and phonon dispersions are computed using scalar relativistic calculations without SOC using the SSSP library v1.0, band structure are also recomputed using fully relativistic calculations with SOC using the PseudoDojo library. Two exceptions are made, namely for monolayer TiNI where PseudoDojo pseudopotentials have been used both for the scalar and fully relativistic calculations, and for monolayer In$_2$ZnS$_4$ where we used the vdW-DF2 functional \cite{lee_df2_09} with C09 exchange (DF2-C09) \cite{cooper_c09_10} for the structural optimization in order to take into account the effect of van der Waals interaction between the ZnS and In$_2$S$_3$ subunits.
 For the materials identified as band insulators, the subsequent fully-relativistic calculations (e.g. to compute the $\mathbb{Z}_2$ invariant) are performed with fixed occupations. A refinement on structural optimization of the QSHIs candidates is performed using a $\mathbf{k}$-point density of 0.1 $\angstrom^{-1}$ and the SSSP precision v1.0 library. \\
The interlayer distance and $E_b$ are computed \cite{mounet_nanotech_18} using two different non-local van-der-Waals functionals: the vdW-DF2 functional \cite{lee_df2_09} with C09 exchange (DF2-C09) \cite{cooper_c09_10} and  the revised Vydrov-Van Voorhis (rVV10) functional \cite{vydrov_vv10_09,sabatini_rvv10_13}.
Topological invariants are computed using Z2pack \cite{soluyanov_z2pack_11,z2pack_gresch_17}. 
Phonons dispersion have been obtained using DFPT \cite{baroni_dfpt_review_01} with the 2D Coulomb cutoff \cite{thibault_prbcutoff_17,thibault_nano_17}, using the SSSP precision library v1.0 and a $\mathbf{q}$-points mesh at least half as dense as the one used for the refinement (so roughly 0.2 $\angstrom^{-1}$).\\
$G_0 W_0$ calculations are perfomed with the Yambo code \cite{yambo_paper_09,yambo_arxiv_2019} on top of DFT-PBE calculations performed with Quantum ESPRESSO. For the Yambo calculations, we use fully relativistic ONCV pseudopotentials from the PseudoDojo library, using the $GW$ version (with complete shell in the valence) \cite{dojo_paper_18} when available. In the $G_0W_0$ calculations we adopt the random integration method (RIM) \cite{yambo_paper_09}, the 2D Coulomb cutoff \cite{yambo_paper_09} and the plasmon-pole approximation for the frequency dependence of the self-energy \cite{godby_prl_1989,godby_prb_1995}. SOC is included self-consistently at the DFT level using spin-orbitals, and fully taken into account at the $G_0W_0$ level using a spinorial Green's function. The inversion strengths introduced in the main text are computed at the $G_0W_0$ level via extrapolation to an infinite dense $\mathbf{k}-$point mesh, by using a fitting function of the form
\begin{equation}
IS(N_k) = \frac{a}{N_k} + \frac{b}{\sqrt{N_k}} + c,
\end{equation}
where $N_k$ is the total number of  $\mathbf{k}-$points in the full Brillouin zone. \\
Part of the calculations are powered by the AiiDA \cite{pizzi_aiida_16} materials' informatics infrastructure.
Direct gap metals (DGM) are identified by computing the direct band gap at every $\mathbf{k}$-point along high-symmetry lines \cite{kpath_86} with a $\mathbf{k}$-point density of 0.01  $\angstrom^{-1}$: if the system is metallic but there is a direct gap at every $\mathbf{k}$-point, i.e.
\begin{equation}
\underset{\mathbf{k}}{\text{min}}\left( \varepsilon_{n_e+1}(\mathbf{k})-\varepsilon_{n_e}(\mathbf{k})\right)>0.01 \text{ eV}
\end{equation}
where $n_e$ is the number of electrons (equal to the number of occupied bands in fully-relativistic calculations with SOC), then the system is considered a DGM.
The magnetic screening is perfomed with collinear DFT-PBE calculations using the procedure employed in Ref.~\cite{mounet_nanotech_18}. First, we explore whether the system is prone to magnetism, by taking the primitive structure and assigning to each atom a collinear random magnetization. Five random configurations are tested by computing the total energy: if at least one configuration results in a magnetic ground state with total energy lower than the one of the initial non-magnetic state, then the configuration is further screened for magnetism otherwise it is considered non magnetic. For the former class of structures, magnetism is further tested by building supercells up to twice the original volume, exploring a range of magnetic, anti-ferromagnetic and ferri-magnetic configurations. Among the different configurations, the one lowest in energy is taken to be the true ground state. In the high-throughput screening for QSHIs we first assume non-magnetic ground states and after, we discard QSHI candidates that are later found to have a magnetic ground state by using the aforementioned protocol.


\begin{thebibliography}{10}
\bibitem{hgte_mol_06}
Markus K{\"o}nig, Steffen Wiedmann, Christoph Br{\"u}ne, Andreas Roth, Hartmut
  Buhmann, Laurens~W. Molenkamp, Xiao-Liang Qi, and Shou-Cheng Zhang.
\newblock Quantum {{Spin Hall Insulator State}} in {{HgTe Quantum Wells}}.
\newblock {\em Science} \textbf{2007}, 318(5851):766--770.

\bibitem{zhang_review_17} Jing Wang and Shou-Cheng Zhang. \newblock Topological states of condensed matter. \newblock {\em Nature Materials}, 16(11):1062--1067 \textbf{2017}.

\bibitem{kane_quantum_2005} C. L. Kane and E. J. Mele, Quantum Spin Hall Effect in Graphene, Phys. Rev. Lett. \textbf{2005}, 95, 226801.
\bibitem{kane_z2_05} C. L. Kane and E. J. Mele, Z2 Topological Order and the Quantum Spin Hall Effect, Phys. Rev. Lett. \textbf{2005}, 95, 146802.

\bibitem{bernevig_strain_06} B. A. Bernevig and S-C. Zhang, Quantum Spin Hall Effect, Phys. Rev. Lett. \textbf{2006}, 96,  106802.
\bibitem{fu_topo3d_07} Liang Fu, C.~L. Kane, and E.~J. Mele.
\newblock Topological {{Insulators}} in {{Three Dimensions}}.
\newblock {\em Physical Review Letters} \textbf{2007}, 98(10):106803.
\bibitem{hasankane_review_10}M. Z. Hasan and C. L. Kane, Colloquium: Topological insulators, Rev. Mod. Phys. \textbf{2010}, 82, 3045.
\bibitem{wte2_sts_17}
Shujie Tang, Chaofan Zhang, Dillon Wong, Zahra Pedramrazi, Hsin-Zon Tsai,
  Chunjing Jia, Brian Moritz, Martin Claassen, Hyejin Ryu, Salman Kahn, Juan
  Jiang, Hao Yan, Makoto Hashimoto, Donghui Lu, Robert~G. Moore, Chan-Cuk
  Hwang, Choongyu Hwang, Zahid Hussain, Yulin Chen, Miguel~M. Ugeda, Zhi Liu,
  Xiaoming Xie, Thomas~P. Devereaux, Michael~F. Crommie, Sung-Kwan Mo, and
  Zhi-Xun Shen.
\newblock Quantum spin {{Hall}} state in monolayer {{1T}}'-{{WTe2}}.
\newblock {\em Nature Physics} \textbf{2017}, 13(7):683--687.

\bibitem{wte2_transp_17}
Zaiyao Fei, Tauno Palomaki, Sanfeng Wu, Wenjin Zhao, Xinghan Cai, Bosong Sun,
  Paul Nguyen, Joseph Finney, Xiaodong Xu, and David~H. Cobden.
\newblock Edge conduction in monolayer {{WTe$_2$}}.
\newblock {\em Nature Physics} \textbf{2017}, 13(7):677--682.

\bibitem{100_science_18}
Sanfeng Wu, Valla Fatemi, Quinn~D. Gibson, Kenji Watanabe, Takashi Taniguchi,
  Robert~J. Cava, and Pablo Jarillo-Herrero.
\newblock Observation of the quantum spin {{Hall}} effect up to 100 kelvin in a
  monolayer crystal.
\newblock {\em Science}  \textbf{2018}, 359(6371):76--79.

\bibitem{reis_bismuthene_17}
F.~Reis, G.~Li, L.~Dudy, M.~Bauernfeind, S.~Glass, W.~Hanke, R.~Thomale,
  J.~Sch{\"a}fer, and R.~Claessen.
\newblock Bismuthene on a {{SiC}} substrate: {{A}} candidate for a
  high-temperature quantum spin {{Hall}} material.
\newblock {\em Science} \textbf{2017}, 357(6348):287--290..

\bibitem{bernevig_book_2013}
B.~Andrei Bernevig and Taylor~L. Hughes.
\newblock {\em Topological Insulators and Topological Superconductors}.
\newblock {Princeton University Press} \textbf{2013}.

\bibitem{qian_quantum_2014}
Xiaofeng Qian, Junwei Liu, Liang Fu, and Ju~Li.
\newblock Quantum spin {{Hall}} effect in two-dimensional transition metal
  dichalcogenides.
\newblock {\em Science} \textbf{2014}, 346(6215):1344--1347.

\bibitem{antonius_temptopo_16}
Gabriel Antonius and Steven~G. Louie.
\newblock Temperature-{{Induced Topological Phase Transitions}}: {{Promoted}}
  versus {{Suppressed Nontrivial Topology}}.
\newblock {\em Physical Review Letters} \textbf{2016}, 117(24):246401.

\bibitem{monserrat_temptopo_16}
Bartomeu Monserrat and David Vanderbilt.
\newblock Temperature {{Effects}} in the {{Band Structure}} of {{Topological
  Insulators}}.
\newblock {\em Physical Review Letters} \textbf{2016}, 117(22):226801.

\bibitem{mounet_nanotech_18}
Nicolas Mounet, Marco Gibertini, Philippe Schwaller, Davide Campi, Andrius
  Merkys, Antimo Marrazzo, Thibault Sohier, Ivano~Eligio Castelli, Andrea
  Cepellotti, Giovanni Pizzi, and Nicola Marzari.
\newblock Two-dimensional materials from high-throughput computational
  exfoliation of experimentally known compounds.
\newblock {\em Nature Nanotechnology} \textbf{2018}, 13(3):246--252.

\bibitem{curtarolo_topoht_12}
Kesong Yang, Wahyu Setyawan, Shidong Wang, Marco~Buongiorno Nardelli, and
  Stefano Curtarolo.
\newblock A search model for topological insulators with high-throughput
  robustness descriptors.
\newblock {\em Nature Materials} \textbf{2012}, 11(7):614--619.

\bibitem{tavazza_spillagearxiv_18}
Kamal Choudhary, Kevin~F. Garrity, and Francesca Tavazza.
\newblock High-throughput discovery of topological materials using spin-orbit
  spillage, Scientific Reports \textbf{2018}, 9, 8534.

\bibitem{Note1} for 165 compounds standard non-magnetic structural optimization did not succeed, mostly due either to the presence of magnetic elements or to unstable structures.
\bibitem{Note2} We further screen out compounds with band inversions (see later in the text) smaller than 20 meV.\bibitem{sgiaro_herma_01}
Claudia Sgiarovello, Maria Peressi, and Raffaele Resta.
\newblock Electron localization in the insulating state: {{Application}} to
  crystalline semiconductors.
\newblock {\em Physical Review B} \textbf{2001}, 64(11):115202.

\bibitem{soluyanov_z2pack_11}
Alexey~A. Soluyanov and David Vanderbilt.
\newblock Computing topological invariants without inversion symmetry.
\newblock {\em Physical Review B} \textbf{2011}, 83(23):235401.

\bibitem{z2pack_gresch_17}
Dominik Gresch, Gabriel Aut{\`e}s, Oleg~V. Yazyev, Matthias Troyer, David
  Vanderbilt, B.~Andrei Bernevig, and Alexey~A. Soluyanov.
\newblock {{Z2Pack}}: {{Numerical}} implementation of hybrid {{Wannier}}
  centers for identifying topological materials.
\newblock {\em Physical Review B} \textbf{2017}, 95(7):075146.

\bibitem{baroni_dfpt_review_01}
Stefano Baroni, Stefano {de Gironcoli}, Andrea Dal~Corso, and Paolo Giannozzi.
\newblock Phonons and related crystal properties from density-functional
  perturbation theory.
\newblock {\em Reviews of Modern Physics} \textbf{2001}, 73(2):515--562.


\bibitem{spillage_vanderbilt_14}
Jianpeng Liu and David Vanderbilt.
\newblock Spin-orbit spillage as a measure of band inversion in insulators.
\newblock {\em Physical Review B} \textbf{2014}, 90(12):125133.



\bibitem{bernevig_topoquantumchem_17}
Barry Bradlyn, L.~Elcoro, Jennifer Cano, M.~G. Vergniory, Zhijun Wang,
  C.~Felser, M.~I. Aroyo, and B.~Andrei Bernevig.
\newblock Topological quantum chemistry.
\newblock {\em Nature}  \textbf{2017}, 547(7663):298--305.

\bibitem{maya_nature_2019}
M.~G. Vergniory, L.~Elcoro, Claudia Felser, Nicolas Regnault, B.~Andrei
  Bernevig, and Zhijun Wang.
\newblock A complete catalogue of high-quality topological materials.
\newblock {\em Nature}  \textbf{2019}, 566(7745):480.

\bibitem{tang_nature_2019}
Feng Tang, Hoi~Chun Po, Ashvin Vishwanath, and Xiangang Wan.
\newblock Comprehensive search for topological materials using symmetry
  indicators.
\newblock {\em Nature}  \textbf{2019}, 566(7745):486.

\bibitem{zhang_nature_2019}
Tiantian Zhang, Yi~Jiang, Zhida Song, He~Huang, Yuqing He, Zhong Fang, Hongming
  Weng, and Chen Fang.
\newblock Catalogue of topological electronic materials.
\newblock {\em Nature}  \textbf{2019}, 566(7745):475.

\bibitem{zunger_beware_2019}
Alex Zunger.
\newblock Beware of plausible predictions of fantasy materials.
\newblock {\em Nature}  \textbf{2019}, 566(7745):447.

\bibitem{klitzing_rev_86}
Klaus {von Klitzing}.
\newblock The quantized {{Hall}} effect.
\newblock {\em Reviews of Modern Physics}  \textbf{1986}, 58(3):519--531.

\bibitem{oleg_bismuth_16}
Gabriel Aut{\`e}s, Anna Isaeva, Luca Moreschini, Jens~C. Johannsen, Andrea
  Pisoni, Ryo Mori, Wentao Zhang, Taisia~G. Filatova, Alexey~N. Kuznetsov,
  L{\'a}szl{\'o} Forr{\'o}, Wouter~Van den Broek, Yeongkwan Kim, Keun~Su Kim,
  Alessandra Lanzara, Jonathan~D. Denlinger, Eli Rotenberg, Aaron Bostwick,
  Marco Grioni, and Oleg~V. Yazyev.
\newblock A novel quasi-one-dimensional topological insulator in bismuth iodide
  $\beta$-{{Bi}}{\textsubscript{4}}{{I}}{\textsubscript{4}}.
\newblock {\em Nature Materials}  \textbf{2016}, 15(2):154--158.

\bibitem{marrazzo_jacutingaite_18}
Antimo Marrazzo, Marco Gibertini, Davide Campi, Nicolas Mounet, and Nicola
  Marzari.
\newblock Prediction of a {{Large}}-{{Gap}} and {{Switchable Kane}}-{{Mele
  Quantum Spin Hall Insulator}}.
\newblock {\em Physical Review Letters}  \textbf{2018}, 120(11):117701.

\bibitem{xenes_review_17}
Alessandro Molle, Joshua Goldberger, Michel Houssa, Yong Xu, Shou-Cheng Zhang,
  and Deji Akinwande.
\newblock Buckled two-dimensional {{Xene}} sheets.
\newblock {\em Nature Materials}  \textbf{2017}, 16(2):163--169.

\bibitem{qian_tellurides_17}
Junwei Liu, Hua Wang, Chen Fang, Liang Fu, and Xiaofeng Qian.
\newblock Van der {{Waals Stacking}}-{{Induced Topological Phase Transition}}
  in {{Layered Ternary Transition Metal Chalcogenides}}.
\newblock {\em Nano Letters}  \textbf{2017}, 17(1):467--475.

\bibitem{monserrat_phsplitting_17}
Bartomeu Monserrat and David Vanderbilt.
\newblock Phonon-assisted spin splitting in centrosymmetric crystals.
\newblock {\em arXiv:1711.06274 [cond-mat]} \textbf{2017}.

\bibitem{zunger_false_2011}
J.~Vidal, X.~Zhang, L.~Yu, J.-W. Luo, and A.~Zunger.
\newblock False-positive and false-negative assignments of topological
  insulators in density functional theory and hybrids.
\newblock {\em Physical Review B}  \textbf{2011}, 84(4):041109.

\bibitem{kpath_86}
Rafael Ram{\'\i}arez and Michael~C. B{\"o}hm.
\newblock Simple geometric generation of special points in brillouin-zone
  integrations. {{Two}}-dimensional bravais lattices.
\newblock {\em International Journal of Quantum Chemistry}  \textbf{1986}, 30(3):391--411.


\bibitem{bernevig_science_06}
B.~Andrei Bernevig, Taylor~L. Hughes, and Shou-Cheng Zhang.
\newblock Quantum {{Spin Hall Effect}} and {{Topological Phase Transition}} in
  {{HgTe Quantum Wells}}.
\newblock {\em Science}  \textbf{2016}, 314(5806):1757--1761.

\bibitem{thygessen_gwconv_13}
Falco H{\"u}ser, Thomas Olsen, and Kristian~S. Thygesen.
\newblock How dielectric screening in two-dimensional crystals affects the
  convergence of excited-state calculations: {{Monolayer MoS}}\$\{\}\_\{2\}\$.
\newblock {\em Physical Review B}  \textbf{2013}, 88(24):245309.

\bibitem{thygessen_2dgw_16}
Filip~A. Rasmussen, Per~S. Schmidt, Kirsten~T. Winther, and Kristian~S.
  Thygesen.
\newblock Efficient many-body calculations for two-dimensional materials using
  exact limits for the screened potential: {{Band}} gaps of
  MoS$_2$,h-BN, and phosphorene.
\newblock {\em Physical Review B}  \textbf{2016}, 94(15):155406.

\bibitem{wang_tini_16}
Aizhu Wang, Zhenhai Wang, Aijun Du, and Mingwen Zhao.
\newblock Band inversion and topological aspects in a {{TiNI}} monolayer.
\newblock {\em Physical Chemistry Chemical Physics}  \textbf{2016}, 18(32):22154--22159.

\bibitem{dresden_htz2_2018}
Xinru Li, Zeying Zhang, Yugui Yao, and Hongbin Zhang.
\newblock High throughput screening for two-dimensional topological insulators.
\newblock {\em 2D Materials}  \textbf{2018}, 5(4):045023.

\bibitem{pd_vyz_2017}
F.~Laufek, A.~Vymazalov{\'a}, and M.~Dr{\'a}bek.
\newblock Powder diffraction study of {{Pd$_2$HgSe$_3$}}.
\newblock {\em Powder Diffraction}  \textbf{2017}, 32(4):244--248.

\bibitem{pdphasediag_vyz_2014}
Milan Dr{\'a}bek, Anna Vymazalov{\'a}, and Franti{\v s}ek Laufek.
\newblock {{The system Hg}}\textendash{}{{Pd}}\textendash{}{{Se at}} 400
  {{\textordmasculine{}C}}: {{phase relations involving tischendorgite and other ternary phases}}.
\newblock {\em The Canadian Mineralogist}  \textbf{2014}, 52(4):763--768.

\bibitem{bismuth2d_prl_06}
Shuichi Murakami.
\newblock Quantum {{Spin Hall Effect}} and {{Enhanced Magnetic Response}} by
  {{Spin}}-{{Orbit Coupling}}.
\newblock {\em Physical Review Letters}  \textbf{2006}, 97(23):236805.

\bibitem{bismuthtopo_nphys_14}
Ilya~K. Drozdov, A.~Alexandradinata, Sangjun Jeon, Stevan Nadj-Perge, Huiwen
  Ji, R.~J. Cava, B.~Andrei~Bernevig, and Ali Yazdani.
\newblock One-dimensional topological edge states of bismuth bilayers.
\newblock {\em Nature Physics}  \textbf{2014}, 10(9):664--669.

\bibitem{bionnb_prb_18}
Lang Peng, Jing-Jing Xian, Peizhe Tang, Angel Rubio, Shou-Cheng Zhang, Wenhao
  Zhang, and Ying-Shuang Fu.
\newblock Visualizing topological edge states of single and double bilayer
  {{Bi}} supported on multibilayer {{Bi}}(111) films.
\newblock {\em Physical Review B}  \textbf{2018}, 98(24):245108.

\bibitem{sabater_prl_13}
C. Sabater, D. Gosálbez-Martínez, J. Fernández-Rossier, J. G. Rodrigo, C. Untiedt, and J. J. Palacios. 
Topologically Protected Quantum Transport in Locally Exfoliated Bismuth at Room Temperature.
\emph{Phys. Rev. Lett.} \textbf{2013}, 110, 176802.

\bibitem{xiang_quantum_2016}
Hui Xiang, Bo~Xu, Jinqiu Liu, Yidong Xia, Haiming Lu, Jiang Yin, and Zhiguo
  Liu.
\newblock Quantum spin {{Hall}} insulator phase in monolayer WTe$_2$ by
  uniaxial strain.
\newblock {\em AIP Advances}  \textbf{2016}, 6(9):095005.

\bibitem{wte_hse_16}
Feipeng Zheng, Chaoyi Cai, Shaofeng Ge, Xuefeng Zhang, Xin Liu, Hong Lu, Yudao
  Zhang, Jun Qiu, Takashi Taniguchi, Kenji Watanabe, Shuang Jia, Jingshan Qi,
  Jian-Hao Chen, Dong Sun, and Ji~Feng.
\newblock On the {{Quantum Spin Hall Gap}} of {{Monolayer 1T}}${'}$-{{WTe$_2$}}.
\newblock {\em Advanced Materials}  \textbf{2016}, 28(24):4845--4851.

\bibitem{lee_df2_09}
Kyuho Lee, {\'E}amonn~D. Murray, Lingzhu Kong, Bengt~I. Lundqvist, and David~C.
  Langreth.
\newblock Higher-accuracy van der {{Waals}} density functional.
\newblock {\em Physical Review B}  \textbf{2010}, 82(8):081101.

\bibitem{cooper_c09_10}
Valentino~R. Cooper.
\newblock Van der {{Waals}} density functional: {{An}} appropriate exchange
  functional.
\newblock {\em Physical Review B}  \textbf{2010}, 81(16):161104.

\bibitem{olsen_z2ht_2019}
Thomas Olsen, Erik Andersen, Takuya Okugawa, Daniele Torelli, Thorsten
  Deilmann, and Kristian~S. Thygesen.
\newblock Discovering two-dimensional topological insulators from
  high-throughput computations.
\newblock {\em Physical Review Materials}  \textbf{2019}, 3(2):024005.

\bibitem{qe_paper_09}
Paolo Giannozzi, Stefano Baroni, Nicola Bonini, Matteo Calandra, Roberto Car,
  Carlo Cavazzoni, {Davide Ceresoli}, Guido~L. Chiarotti, Matteo Cococcioni,
  Ismaila Dabo, Andrea~Dal Corso, Stefano de~Gironcoli, Stefano Fabris, Guido
  Fratesi, Ralph Gebauer, Uwe Gerstmann, Christos Gougoussis, {Anton Kokalj},
  Michele Lazzeri, Layla Martin-Samos, Nicola Marzari, Francesco Mauri,
  Riccardo Mazzarello, {Stefano Paolini}, Alfredo Pasquarello, Lorenzo
  Paulatto, Carlo Sbraccia, Sandro Scandolo, Gabriele Sclauzero, Ari~P.
  Seitsonen, Alexander Smogunov, Paolo Umari, and Renata~M. Wentzcovitch.
\newblock {{QUANTUM ESPRESSO}}: A modular and open-source software project for
  quantum simulations of materials.
\newblock {\em Journal of Physics: Condensed Matter}  \textbf{2009}, 21(39):395502.

\bibitem{qe_paper_17}
P.~Giannozzi, O.~Andreussi, T.~Brumme, O.~Bunau, M.~Buongiorno Nardelli,
  M.~Calandra, R.~Car, C.~Cavazzoni, {D Ceresoli}, M.~Cococcioni, N.~Colonna,
  I.~Carnimeo, A.~Dal Corso, S.~de~Gironcoli, P.~Delugas, R.~A.~DiStasio Jr, {A
  Ferretti}, A.~Floris, G.~Fratesi, G.~Fugallo, R.~Gebauer, U.~Gerstmann,
  F.~Giustino, T.~Gorni, J.~Jia, M.~Kawamura, {H-Y Ko}, A.~Kokalj, E.~K{\"u}{\c
  c}{\"u}kbenli, M.~Lazzeri, M.~Marsili, N.~Marzari, F.~Mauri, N.~L. Nguyen,
  H.-V. Nguyen, {A Otero-de-la-Roza}, L.~Paulatto, S.~Ponc{\'e}, D.~Rocca,
  R.~Sabatini, B.~Santra, M.~Schlipf, A.~P. Seitsonen, A.~Smogunov, {I Timrov},
  T.~Thonhauser, P.~Umari, N.~Vast, X.~Wu, and S.~Baroni.
\newblock Advanced capabilities for materials modelling with {{Q}} uantum
  {{ESPRESSO}}.
\newblock {\em Journal of Physics: Condensed Matter}  \textbf{2017}, 29(46):465901.

\bibitem{sssp_arxiv_18}
Gianluca Prandini, Antimo Marrazzo, Ivano~E. Castelli, Nicolas Mounet, and
  Nicola Marzari, Precision and efficiency in solid-state pseudopotential calculations, \emph{npj Computational Materials} \textbf{2018},
4,72.
\bibitem{hamann_oncv_13}
D.~R. Hamann.
\newblock Optimized norm-conserving {{Vanderbilt}} pseudopotentials.
\newblock {\em Physical Review B} \textbf{2013}, 88(8):085117.

\bibitem{dojo_paper_18}
M.~J. {van Setten}, M.~Giantomassi, E.~Bousquet, M.~J. Verstraete, D.~R.
  Hamann, X.~Gonze, and G.~M. Rignanese.
\newblock The {{PseudoDojo}}: {{Training}} and grading a 85 element optimized
  norm-conserving pseudopotential table.
\newblock {\em Computer Physics Communications} \textbf{2018}, 226:39--54.

\bibitem{cotto_science_16}
Kurt Lejaeghere, Gustav Bihlmayer, Torbj{\"o}rn Bj{\"o}rkman, Peter Blaha,
  Stefan Bl{\"u}gel, Volker Blum, Damien Caliste, Ivano~E. Castelli, Stewart~J.
  Clark, Andrea~Dal Corso, Stefano de~Gironcoli, Thierry Deutsch, John~Kay
  Dewhurst, Igor~Di Marco, Claudia Draxl, Marcin Du{\l}ak, Olle Eriksson,
  Jos{\'e}~A. Flores-Livas, Kevin~F. Garrity, Luigi Genovese, Paolo Giannozzi,
  Matteo Giantomassi, Stefan Goedecker, Xavier Gonze, Oscar Gr{\aa}n{\"a}s,
  E.~K.~U. Gross, Andris Gulans, Fran{\c c}ois Gygi, D.~R. Hamann, Phil~J.
  Hasnip, N.~a.~W. Holzwarth, Diana Iu{\c s}an, Dominik~B. Jochym, Fran{\c
  c}ois Jollet, Daniel Jones, Georg Kresse, Klaus Koepernik, Emine K{\"u}{\c
  c}{\"u}kbenli, Yaroslav~O. Kvashnin, Inka L.~M. Locht, Sven Lubeck, Martijn
  Marsman, Nicola Marzari, Ulrike Nitzsche, Lars Nordstr{\"o}m, Taisuke Ozaki,
  Lorenzo Paulatto, Chris~J. Pickard, Ward Poelmans, Matt I.~J. Probert, Keith
  Refson, Manuel Richter, Gian-Marco Rignanese, Santanu Saha, Matthias
  Scheffler, Martin Schlipf, Karlheinz Schwarz, Sangeeta Sharma, Francesca
  Tavazza, Patrik Thunstr{\"o}m, Alexandre Tkatchenko, Marc Torrent, David
  Vanderbilt, Michiel~J. van Setten, Veronique~Van Speybroeck, John~M. Wills,
  Jonathan~R. Yates, Guo-Xu Zhang, and Stefaan Cottenier.
\newblock Reproducibility in density functional theory calculations of solids.
\newblock {\em Science} \textbf{2016}, 351(6280):aad3000.

\bibitem{mv_smearing_99}
Nicola Marzari, David Vanderbilt, Alessandro De~Vita, and M.~C. Payne.
\newblock Thermal {{Contraction}} and {{Disordering}} of the {{Al}}(110)
  {{Surface}}.
\newblock {\em Physical Review Letters} \textbf{1999}, 82(16):3296--3299.

\bibitem{vydrov_vv10_09}
Oleg~A. Vydrov and Troy Van~Voorhis.
\newblock Nonlocal van der {{Waals Density Functional Made Simple}}.
\newblock {\em Physical Review Letters} \textbf{2009}, 103(6):063004.

\bibitem{sabatini_rvv10_13}
Riccardo Sabatini, Tommaso Gorni, and Stefano {de Gironcoli}.
\newblock Nonlocal van der {{Waals}} density functional made simple and
  efficient.
\newblock {\em Physical Review B} \textbf{2013}, 87(4):041108.

\bibitem{thibault_prbcutoff_17}Thibault Sohier, Matteo Calandra, and Francesco Mauri.
 Density functional perturbation theory for gated two-dimensional heterostructures: Theoretical developments and application to flexural phonons in graphene Phys. Rev. B \textbf{2017}, 96, 075448 
\bibitem{thibault_nano_17}Thibault Sohier, Marco Gibertini, Matteo Calandra, Francesco Mauri, and Nicola Marzari,
Breakdown of Optical Phonons' Splitting in Two-Dimensional Materials,
Nano Letters \textbf{2017}, 17 (6), 3758-3763.

\bibitem{yambo_paper_09}
Andrea Marini, Conor Hogan, Myrta Gr{\"u}ning, and Daniele Varsano.
\newblock Yambo: {{An}} ab initio tool for excited state calculations.
\newblock {\em Computer Physics Communications} \textbf{2009}, 180(8):1392--1403.

\bibitem{yambo_arxiv_2019}
D.~Sangalli, A.~Ferretti, H.~Miranda, C.~Attaccalite, I.~Marri, E.~Cannuccia,
  P.~Melo, M.~Marsili, F.~Paleari, A.~Marrazzo, G.~Prandini, P.~Bonf{\`a},
  M.~O. Atambo, F.~Affinito, M.~Palummo, A.~Molina-S{\'a}nchez, C.~Hogan,
  M.~Gr{\"u}ning, D.~Varsano, and A.~Marini, Many-body perturbation theory calculations using the yambo code,
\emph{ Journal of Physics: Condensed Matter}  \textbf{2019}, 31, 32.
\bibitem{godby_prl_1989}
R.~W. Godby and R.~J. Needs.
\newblock Metal-insulator transition in {{Kohn}}-{{Sham}} theory and
  quasiparticle theory.
\newblock {\em Physical Review Letters} \textbf{2019}, 62(10):1169--1172.

\bibitem{godby_prb_1995}
A.~Oschlies, R.~W. Godby, and R.~J. Needs.
\newblock {{GW}} self-energy calculations of carrier-induced band-gap narrowing
  in n-type silicon.
\newblock {\em Physical Review B} \textbf{1995}, 51(3):1527--1535.

\bibitem{pizzi_aiida_16}
Giovanni Pizzi, Andrea Cepellotti, Riccardo Sabatini, Nicola Marzari, and Boris
  Kozinsky.
\newblock {{AiiDA}}: Automated interactive infrastructure and database for
  computational science.
\newblock {\em Computational Materials Science} \textbf{2016}, 111:218--230.
\end{thebibliography}
\end{document}